\documentclass[aip,
 apl,preprint,
 sd,%
 amsmath,amssymb,
 reprint,%
]{revtex4-1}
\usepackage{graphicx}
\usepackage{color}

\draft 
\begin{document}


\title{Circularly polarized lasing in chiral modulated semiconductor microcavity with GaAs quantum wells}
\author{A. A. Demenev}
\affiliation{Institute of Solid State Physics, Russian Academy of Science, Chernogolovka 142432, Russia}
\author{V. D. Kulakovskii}
\affiliation{Institute of Solid State Physics, Russian Academy of Science, Chernogolovka 142432, Russia}
\author{C.\ Schneider}
\affiliation{Technische Physik and Wilhelm-Conrad-R\"{o}ntgen-Research Center for Complex Material Systems, Universit\"{a}t W\"{u}rzburg, D-97074 W\"{u}rzburg, Am Hubland, Germany}
\author{S.\ Brodbeck}
\affiliation{Technische Physik and Wilhelm-Conrad-R\"{o}ntgen-Research Center for Complex Material Systems, Universit\"{a}t W\"{u}rzburg, D-97074 W\"{u}rzburg, Am Hubland, Germany}
\author{M.\ Kamp}
\affiliation{Technische Physik and Wilhelm-Conrad-R\"{o}ntgen-Research Center for Complex Material Systems, Universit\"{a}t W\"{u}rzburg, D-97074 W\"{u}rzburg, Am Hubland, Germany}
\author{S.\ H\"{o}fling}
\affiliation{Technische Physik and Wilhelm-Conrad-R\"{o}ntgen-Research Center for Complex Material Systems, Universit\"{a}t W\"{u}rzburg, D-97074 W\"{u}rzburg, Am Hubland, Germany}
\author{S.\ V.\ Lobanov}
\affiliation{School of Physics and Astronomy, Cardiff University, Cardiff CF24 3AA, United Kingdom}
\author{T. Weiss}
\affiliation{4$^\mathrm{th}$ Physics Institute and Research Center SCoPE, University of Stuttgart, Stuttgart D-70550, Germany}
\author{N. A. Gippius}
\affiliation{Skolkovo Institute of Science and Technology, Novaya Street 100, Skolkovo 143025, Russia}
\affiliation{A. M. Prokhorov General Physics Institute, Russian Academy of Sciences, Vavilova Street 38, Moscow 119991, Russia}
\author{S. G. Tikhodeev}
\affiliation{A. M. Prokhorov General Physics Institute, Russian Academy of Sciences, Vavilova Street 38, Moscow 119991, Russia}
\affiliation{Institute of Solid State Physics, Russian Academy of Science, Chernogolovka 142432, Russia}
\affiliation{M. V. Lomonosov Moscow State University, Leninskie Gory 1, Moscow 119991, Russia}

\date{27 July 2016}

\begin{abstract}
We report the elliptically, close to circularly polarized  lasing at $\hbar\omega = 1.473$ and 1.522~eV  from
an AlAs/AlGaAs Bragg microcavity with 12 GaAs quantum wells in the active region and
 chiral-etched upper distributed Bragg refractor under optical pump at room  temperature.
The advantage of using the chiral photonic crystal  with a large contrast
of dielectric permittivities is its giant optical activity, allowing to fabricate a very thin half-wave plate, with a thickness of
the order of the emitted light wavelength, and to realize the monolithic control of circular polarization.
\end{abstract}


\maketitle

\begin{figure}[h]
\includegraphics[width=0.9\linewidth]{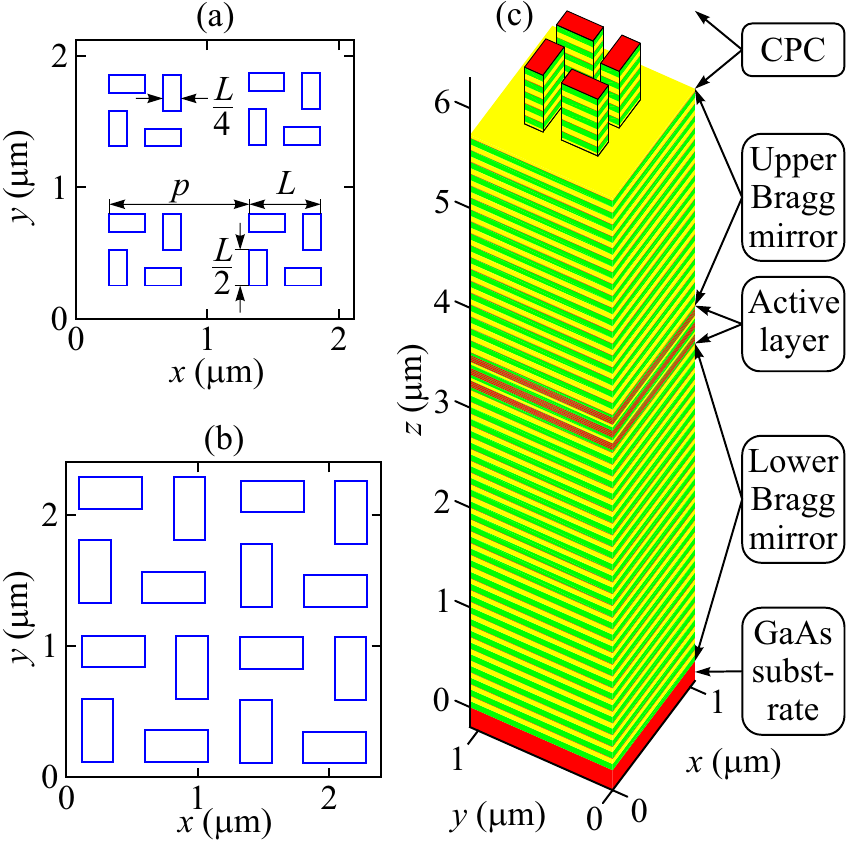}   
\caption{\label{Fig01} (color online). (a,b) The schematics of chiral photonic crystal (CPC) composed of a square lattice of rectangular nanopillars
in Sample~A (panel a) and  B (panel b), $2\times 2$ periods are shown. (c) The schematics of the unit cell
of Sample~A. The nanopillars, composing the CPC,  are etched through the top
4.75 (of 23) Bragg pairs of the upper Bragg mirror of a planar
AlGaAs/AlAs microcavity with twelve GaAs QWs (three groups of four).
Green and yellow colors represent AlAs and AlGaAs $\lambda/4$-layers, respectively. Red color represents GaAs
substrate, smoothing layers between Bragg pairs and active QWs.  A more detailed description of the cavity design
is given in the text and Supplementary material.  }
\end{figure}

\begin{figure}[h]
\includegraphics[width=0.98\linewidth]{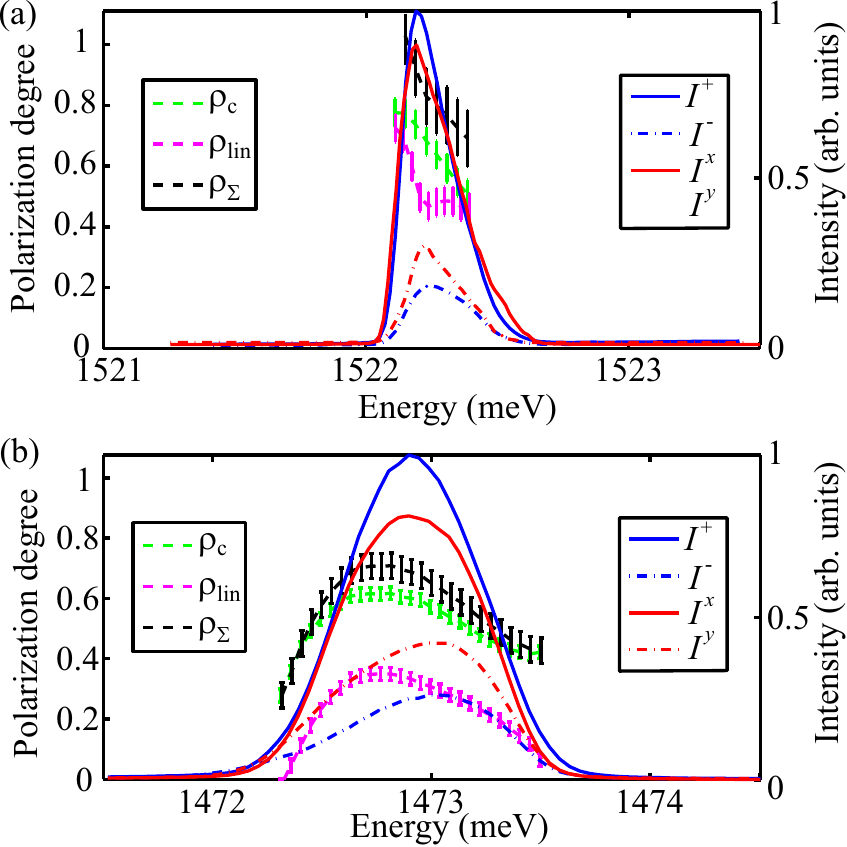}  \\ %
\caption{\label{Fig02}  (color online). The emission intensity spectra of Sample~A (a) and Sample B (b) at $T=300$~K at pump intensity above the threshold in circular $I^{\pm}$ (blue solid and
dash-dotted lines) and linear $I^{x,y}$ (red lines) polarizations. Green, magenta, and black lines with errorbars show the circular,  linear, and
total polarization degrees, $\rho_c$,  $\rho_\mathrm{lin}$, $\rho_\Sigma = \sqrt{\rho_c^2+\rho_\mathrm{lin}^2}$, respectively.
}%
\end{figure}

\begin{figure}[h]
\includegraphics[width=0.98\linewidth]{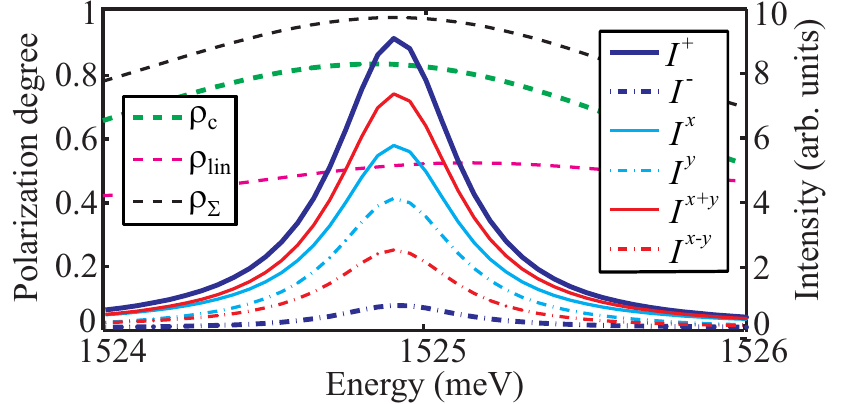}  \\  
\caption{\label{Fig03}  (color online). Calculated (for Sample A) emission intensity spectra  in right- and left-circular polarizations (thick blue solid and dash-dotted lines). The corresponding
circular polarization degree spectra is shown by thick dashed green line. Cyan and red lines show the linearly polarized (in $xy$ and diagonal directions respectively) intensities
$I^x, I^y,I^{x+y},I^{x-y}$, calculated assuming that all oscillating dipoles are aligned along $x$ direction. The
resulting linear and total polarization degrees $\rho_\mathrm{lin}$ and $\rho_\Sigma$ are shown as magenta and black dashed lines.}
\end{figure}

Modern nanofabrication technologies allow to
realize  photonic structures --- photonic crystals and metamaterials --- with extraordinary
optical properties~\cite{Yablonovitch1987,Fan1997,Miyai2006}.
In particular, chiral photonic structures are known to demonstrate a giant optical activity,
several orders of magnitude stronger than natural materials~\cite{Papakostas2003,Kuwata-Gonokami2005,Kwon2008,Liu2009,Hentschel2012,Yin2013}.
Recently, it has been demonstrated that incorporating a chiral photonic structure into
a planar GaAs waveguide or a semiconductor microcavity (MC) with embedded light-emitting achiral InAs quantum dots (QD)
allows to achieve highly circularly polarized light emission,  without applying magnetic field and
without the need of thick quarter-waveplates~\cite{Maksimov2014,Lobanov2015,Lobanov2015a}. The effect is due to
the modification of the symmetry and density of environmentally
allowed electromagnetic modes relative to that in free
space due to the chiral nanostructuring, which, in  turn,  affects the spontaneous emission rate, directional pattern,
and polarization~\cite{Konishi2011,Shitrit2013}.
This method has considerable advantages: small size, very simple operation, and compatibility with semiconductor fabrication process.
In this Letter we demonstrate that the method works for the stimulated emission as well, and demonstrate a highly circularly
polarized  lasing from an AlGaAs/AlAs microcavity with chirally etched top Bragg mirror with GaAs quantum wells (QW) in the active
cavity. To the best of our knowledge,  previously the elliptically polarized lasing  with a good degree of circular polarization
with monolithic control of circular polarization was realized only on a quantum cascade laser in the THz range of frequencies~\cite{Rauter2014}.

A chiral photonic crystal (CPC)  is fabricated from the AlAs/AlGaAs/GaAs high Q-factor MC grown by molecular beam epitaxy on a (001)-oriented GaAs.
The full planar cavity consists of  a lower and an upper Bragg reflectors with 27 and 23
 pairs of AlAs/Al$_{0.20}$Ga$_{0.80}$As layers, respectively,  with 3~nm GaAs smoothing
layer after each pair in the Bragg reflectors and an active layer with three groups of four 13~nm GaAs QWs separated by
4~nm AlAs bariers. The nominal thicknesses of the AlAs and Al$_{0.20}$Ga$_{0.80}$As
layers  in Bragg mirrors are (68 $\pm$ 3)~nm and (58 $\pm$ 3)~nm, respectively.
The Bragg pairs are deposited on the wafer
with a slight wedge from the center to the circumference, resulting in a blueshift of the cavity resonance
which can amount up to  $\approx$200~meV..
 It consists of the central group of four GaAs QWs with three AlAs barriers between them,
surrounded by 32~nm AlAs and 26~nm Al$_{0.20}$Ga$_{0.80}$As layers, and symmetric siding groups of four AlAs/GaAs barier/QW layers with 28~nm AlAs
trailing layers.\footnote{See a more detailed description of the cavity in the Supplementary material.}

\begin{figure*}[t]
\includegraphics[width=0.95\linewidth]{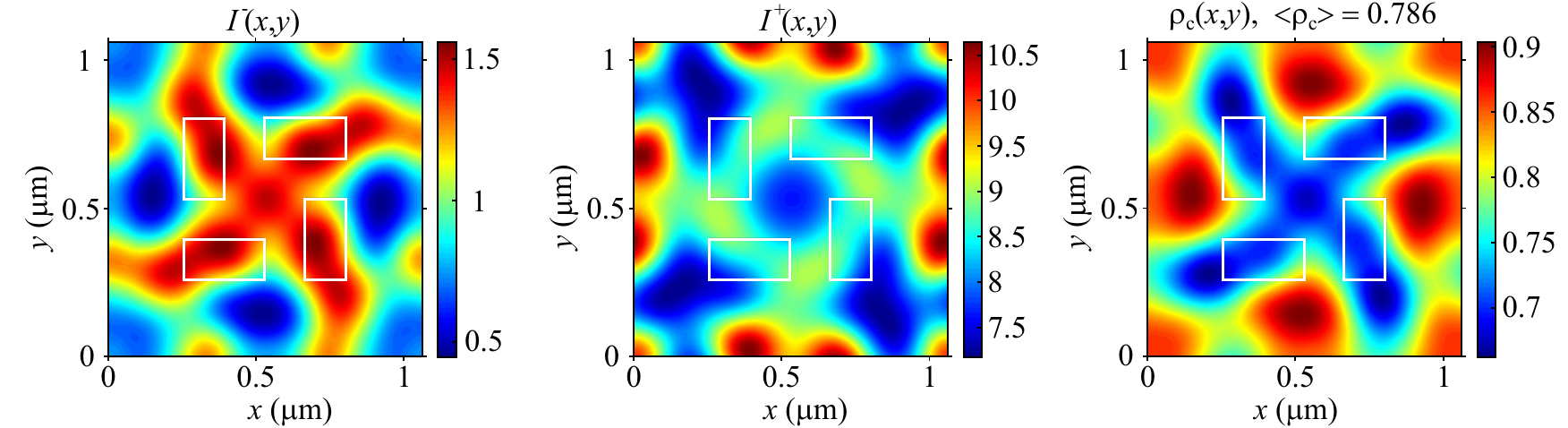}  \\  
\caption{\label{Fig04}  (color online). Calculated spatial emission intensity
distributions in Sample~A in left ($I^-(x,y)$) and right ($I^+(x,y)$) circular
polarizations (left and central panels), and spatial
distribution of circular polarization degree $\rho_c(x,y)$ (right panel).}%
\end{figure*}

A chiral layer is fabricated by nanolithography and dry etching through top 4.75 Bragg pairs of the upper mirror
(which means through the four top Bragg pairs, the AlAs layer and 1/2 of the Al$_{0.20}$Ga$_{0.80}$As
layer of the 5th Bragg pair).
The schematics of the CPC is depicted in Fig.~1.
It consists of a square lattice of rectangular pillars that have a broken in-plane mirror symmetry but possess a fourfold
rotational axis and possess the potential to demonstrate a strong optical activity~\cite{Konishi2008}.
The vertical walls of nanopillars are normal to the $[110]$ and $[\bar{1}10]$ crystallographic
directions. This structure has a C$_4$ point symmetry, and it is three-dimentionally chiral, because it does not have planes of mirror symmetry,
including the horizontal one~\cite{Kwon2008}.

 Two different periodic structures (Sample~A and B hereafter) have
been fabricated, basing on the theoretical calculations explained below and shown schematically in Figs.~1a and 1b. Sample~A (panel a)
has the period $p=1060$~nm and pillar feature size $L=544$~nm, and Sample~B (panel b) has  $p=1200$~nm and $L=960$~nm. The horizontal size
of chiral structure in each sample is approximately $50\times50\, \mu\mathrm{m}^2$. The Samples A and B are cleaved from different horizontal parts of
the wafer, and show different photon energies of the main MC mode, around 1523 and 1473~meV at room temperature, respectively.

The samples are held at room temperature. The excitation is carried out with a Ti-sapphire laser in the spectral range
of the first reflection minimum of the MC. The laser spot has
a diameter of about $10~\mu$m. The emission is collected in an angle range of $\pm 15^\circ$. It is dispersed by a monochromator and
 detected by a Si CCD camera. The polarization of the
luminescence is analyzed by a quarter wave retarder and linear polarizers.

The emission intensity  from both samples at low pump intensities depends weakly on the angle, its spectral width reaches
a few meV. With increase in the excitation power the emission spectrum and intensity show a threshold-like transition to lasing regime
at $P=P_\mathrm{thr}$. Above the threshold the emission line becomes narrow. The full width at half maximum is about 0.23 meV
at $P=1.1 P_\mathrm{thr}$ and increases to $\sim 0.5$ meV at $P=2.5 P_\mathrm{thr}$. Figures~2a and b display the measured
emission intensities (from Samples~A and B, respectively)  in two circular ($I^+, I^-$) and two linear ($I^x, I^y$) polarizations as functions of photon energy $\hbar \omega$,
at the normal to the MC plane at  $P \sim 2P_\mathrm{thr}$ and zero magnetic field.
(The emission intensities in two diagonal linear polarizations $x\pm y$ are not shown as they differ very weakly from each other.)
The dashed lines in Figs. 2a and b show the degrees of circular, linear, and
total polarization defined as
\begin{eqnarray}
 \rho_c & = & \frac{I^+ - I^-}{ I^+ + I^-},  \rho_{\mathrm{lin},x} =\frac{I^x - I^y}{ I^x + I^y},   \rho_{\mathrm{lin},d}  =\frac{I^{x+y} - I^{x-y}}{ I^{x+y} +I^{x-y}},
  \nonumber \\ \label{Eq:rhoCandXY}
\rho_\Sigma & = &  \sqrt{\rho_c^2+\rho_{\mathrm{lin},x}^2+\rho_{\mathrm{lin},d}^2}.
\end{eqnarray}

It is seen that in both samples the lasing is nearly completely elliptically polarized,
with polarization degrees at the intensity maxima as large as
 $\rho_c \sim 80\% $,  $\rho_{\mathrm{lin},x}  \sim 50\% $,  and $\rho_\Sigma \sim 95\%$ in Sample~A (Fig 2a),
and slightly smaller  in Sample~B (Fig 2b), $\rho_c \sim 60\% $,  $\rho_{xy}  \sim 35\% $, and  $\rho_\Sigma \sim 80\%$.

\begin{figure}
\includegraphics[width=0.99\linewidth]{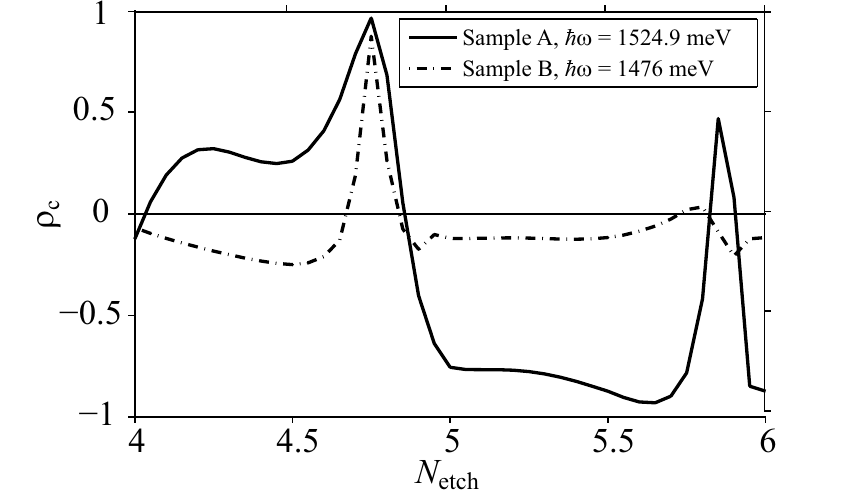}  \\  
\caption{\label{Fig05} Calculated dependencies of maximum circular polarization degree $\rho_c$ on etching depth $N_\mathrm{etch}$ (measured as the number of etched Bragg pairs) for
Sample A (solid line) and Sample B (dash-dotted line) structures. }%
\end{figure}

The reciprocity and symmetry analysis of the structure shows that the CPC in the structures works as a
waveplate, exploring the Fabry-Perot interference between the vertically propagating modes in the slab, which allows reaching
nearly a 100\% circular polarization of the transmission~\cite{Lobanov2015}. To optimize the chiral structures for obtaining
a high  degree of circular polarization degree (DCP) of light emission we have calculated the frequency dependence of emission in right and left
circular polarizations, using the optical scattering matrix and Fourier modal method~\cite{Whittaker1999,Maksimov2014,Lobanov2015,Lobanov2015a}. In this approximation
the emission is calculated actually for  homogeneously distributed oscillating point dipoles  in the QW plane, which are driven by external excitation
and emit incoherently, so that the intensities rather than the electromagnetic fields are summed up at the receiver. This approximation
(so called weak coupling limit) is not completely valid for describing the lasing. But below and slightly above  the threshold of the
lasing regime
it might be a reasonable starting point for optimization of the structures.

 In this approximation, assuming the overall C$_4$ symmetry
of the system, only a circular polarization of emission can be expected. The linear polarization is absent, because the oscillating
dipoles are assumed to be randomly linearly polarized in $xy$ plane. The calculated emission intensities in right and left circular
polarizations $I^\pm$ for a structure with parameters of Sample~A are shown in Fig.~3 as blue solid and dash-dotted lines. The intensities
are normalized to the emission intensity of the same oscillating dipoles in vacuum.
The resulting dependence of $\rho_c(\hbar\omega)$ is  shown in Fig.~3 by the dashed green line. It can be seen that the emission is expected to
be strongly circularly polarized, with $\rho_c$ up to 80\%, in agreement with the experiment.

Additional insight into the mechanism of circularly polarized emission is provided in Fig.~4 with the calculated spatial
distribution of the circularly polarized emission intensity in Sample~A at the resonant
frequency $\hbar \omega = 1.522$~eV, as well as the spatial distribution of the DCP of emission $\rho_c(x,y)$, over the CPC unit cell.

Note that $\rho_ c$ is very sensitive to the etching depth. Figure~5 shows the calculated dependence of the maximum $\rho_c$  of
emission from Samples~A (solid line) and B (dash-dotted line) as functions of etching
depth $N_\mathrm{etch}$ (measured in the number of etched Bragg pairs). Thus, the  structures  employed in our experiments with
$N_\mathrm{etch} \sim 4.75$ are in agreement with these calculations, corresponding to the maxima of the expected $\rho_ c$.

Figure 2 shows that the measured emission in the lasing mode  shows as well a  pronounced linear polarization. The linear polarization indicates that the system
has a lowered lateral symmetry. This might be due to a lower symmetry of the fabricated CPC structures, and/or due to the stimulated
alignment of the oscillating dipoles in the lasing regime. The calculations show that the degree of circular polarization does not
change with the alignment, but  the linear polarization appears. This is illustrated in Fig.~3 where cyan  and red lines show the
calculated emission intensities $I^{x,y}, I^{x\pm y}$ in $xy$ and diagonal linear polarizations, assuming that all oscillating dipoles in the model are aligned
along $x$ axis. The  linear polarization degree $\rho_\mathrm{lin}=\sqrt{\rho_{\mathrm{lin},x}^2+\rho_{\mathrm{lin},d}^2}$ and total
polarization degree $ \rho_\Sigma = \sqrt{\rho_c^2+\rho_\mathrm{lin}^2}$ are shown as magenta and black dashed lines in Fig.~3.
 It is seen that the alignment of the oscillating dipoles leads to a pronounced linear polarization. The deviations from the C$_4$ symmetry cause
 the linear polarization of emission as well, see in Supplementary material. However, the DCP in this case is usually less than in the perfect structure.

To conclude, we have found that the nanotechnology allows fabricating chiral photonic crystals with light emitting
 GaAs quantum wells inside a planar MC to realize lasing
 with a high circular polarization of the light  emission in the absence of a magnetic field. The advantage of using the CPCs with a large contrast
of dielectric permittivities is its giant optical activity. This allows one to fabricate a very thin ``waveplate'', with a thickness of
 the order of the emitted light wavelength.  One more advantage of CPC half-wave plates lies in the fact that they, unlike the traditional
 ones, have in-plane rotational isotropy due to the C$_4$ symmetry.

Acknowledgement. This work has been funded by Russian Scientific Foundation (grant 14-12-01372)
 and State of Bavaria.
 We are grateful to K. Konishi, L. Kuipers, M. Kuwata-Gonokami, R. Oulton, H. Tamaru, and F. Capasso for fruitful discussions, and
 M. Emmerling for preparing the nanopillars.


\begin{thebibliography}{17}%
\makeatletter
\providecommand \@ifxundefined [1]{%
 \@ifx{#1\undefined}
}%
\providecommand \@ifnum [1]{%
 \ifnum #1\expandafter \@firstoftwo
 \else \expandafter \@secondoftwo
 \fi
}%
\providecommand \@ifx [1]{%
 \ifx #1\expandafter \@firstoftwo
 \else \expandafter \@secondoftwo
 \fi
}%
\providecommand \natexlab [1]{#1}%
\providecommand \enquote  [1]{``#1''}%
\providecommand \bibnamefont  [1]{#1}%
\providecommand \bibfnamefont [1]{#1}%
\providecommand \citenamefont [1]{#1}%
\providecommand \href@noop [0]{\@secondoftwo}%
\providecommand \href [0]{\begingroup \@sanitize@url \@href}%
\providecommand \@href[1]{\@@startlink{#1}\@@href}%
\providecommand \@@href[1]{\endgroup#1\@@endlink}%
\providecommand \@sanitize@url [0]{\catcode `\\12\catcode `\$12\catcode
  `\&12\catcode `\#12\catcode `\^12\catcode `\_12\catcode `\%12\relax}%
\providecommand \@@startlink[1]{}%
\providecommand \@@endlink[0]{}%
\providecommand \url  [0]{\begingroup\@sanitize@url \@url }%
\providecommand \@url [1]{\endgroup\@href {#1}{\urlprefix }}%
\providecommand \urlprefix  [0]{URL }%
\providecommand \Eprint [0]{\href }%
\providecommand \doibase [0]{http://dx.doi.org/}%
\providecommand \selectlanguage [0]{\@gobble}%
\providecommand \bibinfo  [0]{\@secondoftwo}%
\providecommand \bibfield  [0]{\@secondoftwo}%
\providecommand \translation [1]{[#1]}%
\providecommand \BibitemOpen [0]{}%
\providecommand \bibitemStop [0]{}%
\providecommand \bibitemNoStop [0]{.\EOS\space}%
\providecommand \EOS [0]{\spacefactor3000\relax}%
\providecommand \BibitemShut  [1]{\csname bibitem#1\endcsname}%
\let\auto@bib@innerbib\@empty
\bibitem [{\citenamefont {Yablonovitch}(1987)}]{Yablonovitch1987}%
  \BibitemOpen
  \bibfield  {author} {\bibinfo {author} {\bibfnamefont {E.}~\bibnamefont
  {Yablonovitch}},\ }\href@noop {} {\bibfield  {journal} {\bibinfo  {journal}
  {Phys.\ Rev.\ Lett.}\ }\textbf {\bibinfo {volume} {58}},\ \bibinfo {pages}
  {2059} (\bibinfo {year} {1987})}\BibitemShut {NoStop}%
\bibitem [{\citenamefont {Fan}\ \emph {et~al.}(1997)\citenamefont {Fan},
  \citenamefont {Villeneuve}, \citenamefont {Joannopoulos},\ and\ \citenamefont
  {Schubert}}]{Fan1997}%
  \BibitemOpen
  \bibfield  {author} {\bibinfo {author} {\bibfnamefont {S.}~\bibnamefont
  {Fan}}, \bibinfo {author} {\bibfnamefont {P.~R.}\ \bibnamefont {Villeneuve}},
  \bibinfo {author} {\bibfnamefont {J.}~\bibnamefont {Joannopoulos}}, \ and\
  \bibinfo {author} {\bibfnamefont {E.~F.}\ \bibnamefont {Schubert}},\ }\href
  {\doibase 10.1103/PhysRevLett.78.3294} {\bibfield  {journal} {\bibinfo
  {journal} {Phys.\ Rev.\ Lett.}\ }\textbf {\bibinfo {volume} {78}},\ \bibinfo
  {pages} {3294} (\bibinfo {year} {1997})}\BibitemShut {NoStop}%
\bibitem [{\citenamefont {Miyai}\ \emph {et~al.}(2006)\citenamefont {Miyai},
  \citenamefont {Sakai}, \citenamefont {Okano}, \citenamefont {Kunishi},
  \citenamefont {Ohnishi},\ and\ \citenamefont {Noda}}]{Miyai2006}%
  \BibitemOpen
  \bibfield  {author} {\bibinfo {author} {\bibfnamefont {E.}~\bibnamefont
  {Miyai}}, \bibinfo {author} {\bibfnamefont {K.}~\bibnamefont {Sakai}},
  \bibinfo {author} {\bibfnamefont {T.}~\bibnamefont {Okano}}, \bibinfo
  {author} {\bibfnamefont {W.}~\bibnamefont {Kunishi}}, \bibinfo {author}
  {\bibfnamefont {D.}~\bibnamefont {Ohnishi}}, \ and\ \bibinfo {author}
  {\bibfnamefont {S.}~\bibnamefont {Noda}},\ }\href {\doibase 10.1038/441946a}
  {\bibfield  {journal} {\bibinfo  {journal} {Nature}\ }\textbf {\bibinfo
  {volume} {441}},\ \bibinfo {pages} {946} (\bibinfo {year}
  {2006})}\BibitemShut {NoStop}%
\bibitem [{\citenamefont {Papakostas}\ \emph {et~al.}(2003)\citenamefont
  {Papakostas}, \citenamefont {Potts}, \citenamefont {Bagnall}, \citenamefont
  {Prosvirnin}, \citenamefont {Coles},\ and\ \citenamefont
  {Zheludev}}]{Papakostas2003}%
  \BibitemOpen
  \bibfield  {author} {\bibinfo {author} {\bibfnamefont {A.}~\bibnamefont
  {Papakostas}}, \bibinfo {author} {\bibfnamefont {A.}~\bibnamefont {Potts}},
  \bibinfo {author} {\bibfnamefont {D.~M.}\ \bibnamefont {Bagnall}}, \bibinfo
  {author} {\bibfnamefont {S.~L.}\ \bibnamefont {Prosvirnin}}, \bibinfo
  {author} {\bibfnamefont {H.~J.}\ \bibnamefont {Coles}}, \ and\ \bibinfo
  {author} {\bibfnamefont {N.~I.}\ \bibnamefont {Zheludev}},\ }\href {\doibase
  10.1103/PhysRevLett.90.107404} {\bibfield  {journal} {\bibinfo  {journal}
  {Phys. Rev. Lett.}\ }\textbf {\bibinfo {volume} {90}},\ \bibinfo {pages}
  {107404} (\bibinfo {year} {2003})}\BibitemShut {NoStop}%
\bibitem [{\citenamefont {Kuwata-Gonokami}\ \emph {et~al.}(2005)\citenamefont
  {Kuwata-Gonokami}, \citenamefont {Saito}, \citenamefont {Ino}, \citenamefont
  {Kauranen}, \citenamefont {Jefimovs}, \citenamefont {Vallius}, \citenamefont
  {Turunen},\ and\ \citenamefont {Svirko}}]{Kuwata-Gonokami2005}%
  \BibitemOpen
  \bibfield  {author} {\bibinfo {author} {\bibfnamefont {M.}~\bibnamefont
  {Kuwata-Gonokami}}, \bibinfo {author} {\bibfnamefont {N.}~\bibnamefont
  {Saito}}, \bibinfo {author} {\bibfnamefont {Y.}~\bibnamefont {Ino}}, \bibinfo
  {author} {\bibfnamefont {M.}~\bibnamefont {Kauranen}}, \bibinfo {author}
  {\bibfnamefont {K.}~\bibnamefont {Jefimovs}}, \bibinfo {author}
  {\bibfnamefont {T.}~\bibnamefont {Vallius}}, \bibinfo {author} {\bibfnamefont
  {J.}~\bibnamefont {Turunen}}, \ and\ \bibinfo {author} {\bibfnamefont
  {Y.}~\bibnamefont {Svirko}},\ }\href {\doibase 10.1103/PhysRevLett.95.227401}
  {\bibfield  {journal} {\bibinfo  {journal} {Phys. Rev. Lett.}\ }\textbf
  {\bibinfo {volume} {95}},\ \bibinfo {eid} {227401} (\bibinfo {year}
  {2005})}\BibitemShut {NoStop}%
\bibitem [{\citenamefont {Kwon}, \citenamefont {Werner},\ and\ \citenamefont
  {Werner}(2008)}]{Kwon2008}%
  \BibitemOpen
  \bibfield  {author} {\bibinfo {author} {\bibfnamefont {D.-H.}\ \bibnamefont
  {Kwon}}, \bibinfo {author} {\bibfnamefont {P.~L.}\ \bibnamefont {Werner}}, \
  and\ \bibinfo {author} {\bibfnamefont {D.~H.}\ \bibnamefont {Werner}},\
  }\href {\doibase doi:10.1364/OE.16.011802} {\bibfield  {journal} {\bibinfo
  {journal} {Opt. Express}\ }\textbf {\bibinfo {volume} {16}},\ \bibinfo
  {pages} {11802} (\bibinfo {year} {2008})},\ \bibinfo {note} {d.-H. Kwon P. L.
  Werner D. H. Werner}\BibitemShut {NoStop}%
\bibitem [{\citenamefont {Liu}\ \emph {et~al.}(2009)\citenamefont {Liu},
  \citenamefont {Liu}, \citenamefont {Zhu},\ and\ \citenamefont
  {Giessen}}]{Liu2009}%
  \BibitemOpen
  \bibfield  {author} {\bibinfo {author} {\bibfnamefont {N.}~\bibnamefont
  {Liu}}, \bibinfo {author} {\bibfnamefont {H.}~\bibnamefont {Liu}}, \bibinfo
  {author} {\bibfnamefont {S.}~\bibnamefont {Zhu}}, \ and\ \bibinfo {author}
  {\bibfnamefont {H.}~\bibnamefont {Giessen}},\ }\href
  {http://dx.doi.org/10.1038/nphoton.2009.4} {\bibfield  {journal} {\bibinfo
  {journal} {Nat. Photon.}\ }\textbf {\bibinfo {volume} {3}},\ \bibinfo {pages}
  {157} (\bibinfo {year} {2009})}\BibitemShut {NoStop}%
\bibitem [{\citenamefont {Hentschel}\ \emph {et~al.}(2012)\citenamefont
  {Hentschel}, \citenamefont {Wu}, \citenamefont {Sch\"{a}ferling},
  \citenamefont {Bai}, \citenamefont {Li},\ and\ \citenamefont
  {Giessen}}]{Hentschel2012}%
  \BibitemOpen
  \bibfield  {author} {\bibinfo {author} {\bibfnamefont {M.}~\bibnamefont
  {Hentschel}}, \bibinfo {author} {\bibfnamefont {L.}~\bibnamefont {Wu}},
  \bibinfo {author} {\bibfnamefont {M.}~\bibnamefont {Sch\"{a}ferling}},
  \bibinfo {author} {\bibfnamefont {P.}~\bibnamefont {Bai}}, \bibinfo {author}
  {\bibfnamefont {E.~P.}\ \bibnamefont {Li}}, \ and\ \bibinfo {author}
  {\bibfnamefont {H.}~\bibnamefont {Giessen}},\ }\href {\doibase
  10.1021/nn304283y} {\bibfield  {journal} {\bibinfo  {journal} {ACS Nano}\
  }\textbf {\bibinfo {volume} {6}},\ \bibinfo {pages} {10355} (\bibinfo {year}
  {2012})}.
 \BibitemShut {NoStop}%
\bibitem [{\citenamefont {Yin}\ \emph {et~al.}(2013)\citenamefont {Yin},
  \citenamefont {Sch\"{a}ferling}, \citenamefont {Metzger},\ and\ \citenamefont
  {Giessen}}]{Yin2013}%
  \BibitemOpen
  \bibfield  {author} {\bibinfo {author} {\bibfnamefont {X.}~\bibnamefont
  {Yin}}, \bibinfo {author} {\bibfnamefont {M.}~\bibnamefont
  {Sch\"{a}ferling}}, \bibinfo {author} {\bibfnamefont {B.}~\bibnamefont
  {Metzger}}, \ and\ \bibinfo {author} {\bibfnamefont {H.}~\bibnamefont
  {Giessen}},\ }\href {\doibase 10.1021/nl403705k} {\bibfield  {journal}
  {\bibinfo  {journal} {Nano Letters}\ }\textbf {\bibinfo {volume} {13}},\
  \bibinfo {pages} {6238} (\bibinfo {year} {2013})}.\
\BibitemShut {NoStop}%
\bibitem [{\citenamefont {Maksimov}\ \emph {et~al.}(2014)\citenamefont
  {Maksimov}, \citenamefont {Tartakovskii}, \citenamefont {Filatov},
  \citenamefont {Lobanov}, \citenamefont {Gippius}, \citenamefont {Tikhodeev},
  \citenamefont {Schneider}, \citenamefont {Kamp}, \citenamefont {Maier},
  \citenamefont {H\"ofling},\ and\ \citenamefont {Kulakovskii}}]{Maksimov2014}%
  \BibitemOpen
  \bibfield  {author} {\bibinfo {author} {\bibfnamefont {A.~A.}\ \bibnamefont
  {Maksimov}}, \bibinfo {author} {\bibfnamefont {I.~I.}\ \bibnamefont
  {Tartakovskii}}, \bibinfo {author} {\bibfnamefont {E.~V.}\ \bibnamefont
  {Filatov}}, \bibinfo {author} {\bibfnamefont {S.~V.}\ \bibnamefont
  {Lobanov}}, \bibinfo {author} {\bibfnamefont {N.~A.}\ \bibnamefont
  {Gippius}}, \bibinfo {author} {\bibfnamefont {S.~G.}\ \bibnamefont
  {Tikhodeev}}, \bibinfo {author} {\bibfnamefont {C.}~\bibnamefont
  {Schneider}}, \bibinfo {author} {\bibfnamefont {M.}~\bibnamefont {Kamp}},
  \bibinfo {author} {\bibfnamefont {S.}~\bibnamefont {Maier}}, \bibinfo
  {author} {\bibfnamefont {S.}~\bibnamefont {H\"ofling}}, \ and\ \bibinfo
  {author} {\bibfnamefont {V.~D.}\ \bibnamefont {Kulakovskii}},\ }\href
  {\doibase 10.1103/PhysRevB.89.045316} {\bibfield  {journal} {\bibinfo
  {journal} {Phys. Rev. B}\ }\textbf {\bibinfo {volume} {89}},\ \bibinfo
  {pages} {045316} (\bibinfo {year} {2014})}\BibitemShut {NoStop}%
\bibitem [{\citenamefont {Lobanov}\ \emph
  {et~al.}(2015{\natexlab{a}})\citenamefont {Lobanov}, \citenamefont {Weiss},
  \citenamefont {Gippius}, \citenamefont {Tikhodeev}, \citenamefont
  {Kulakovskii}, \citenamefont {Konishi},\ and\ \citenamefont
  {Kuwata-Gonokami}}]{Lobanov2015}%
  \BibitemOpen
  \bibfield  {author} {\bibinfo {author} {\bibfnamefont {S.~V.}\ \bibnamefont
  {Lobanov}}, \bibinfo {author} {\bibfnamefont {T.}~\bibnamefont {Weiss}},
  \bibinfo {author} {\bibfnamefont {N.~A.}\ \bibnamefont {Gippius}}, \bibinfo
  {author} {\bibfnamefont {S.~G.}\ \bibnamefont {Tikhodeev}}, \bibinfo {author}
  {\bibfnamefont {V.~D.}\ \bibnamefont {Kulakovskii}}, \bibinfo {author}
  {\bibfnamefont {K.}~\bibnamefont {Konishi}}, \ and\ \bibinfo {author}
  {\bibfnamefont {M.}~\bibnamefont {Kuwata-Gonokami}},\ }\href {\doibase
  10.1364/OL.40.001528} {\bibfield  {journal} {\bibinfo  {journal} {Opt.
  Lett.}\ }\textbf {\bibinfo {volume} {40}},\ \bibinfo {pages} {1528} (\bibinfo
  {year} {2015}{\natexlab{a}})}\BibitemShut {NoStop}%
\bibitem [{\citenamefont {Lobanov}\ \emph
  {et~al.}(2015{\natexlab{b}})\citenamefont {Lobanov}, \citenamefont
  {Tikhodeev}, \citenamefont {Gippius}, \citenamefont {Maksimov}, \citenamefont
  {Filatov}, \citenamefont {Tartakovskii}, \citenamefont {Kulakovskii},
  \citenamefont {Weiss}, \citenamefont {Schneider}, \citenamefont {Ge\ss{}ler},
  \citenamefont {Kamp},\ and\ \citenamefont {H\"ofling}}]{Lobanov2015a}%
  \BibitemOpen
  \bibfield  {author} {\bibinfo {author} {\bibfnamefont {S.~V.}\ \bibnamefont
  {Lobanov}}, \bibinfo {author} {\bibfnamefont {S.~G.}\ \bibnamefont
  {Tikhodeev}}, \bibinfo {author} {\bibfnamefont {N.~A.}\ \bibnamefont
  {Gippius}}, \bibinfo {author} {\bibfnamefont {A.~A.}\ \bibnamefont
  {Maksimov}}, \bibinfo {author} {\bibfnamefont {E.~V.}\ \bibnamefont
  {Filatov}}, \bibinfo {author} {\bibfnamefont {I.~I.}\ \bibnamefont
  {Tartakovskii}}, \bibinfo {author} {\bibfnamefont {V.~D.}\ \bibnamefont
  {Kulakovskii}}, \bibinfo {author} {\bibfnamefont {T.}~\bibnamefont {Weiss}},
  \bibinfo {author} {\bibfnamefont {C.}~\bibnamefont {Schneider}}, \bibinfo
  {author} {\bibfnamefont {J.}~\bibnamefont {Ge\ss{}ler}}, \bibinfo {author}
  {\bibfnamefont {M.}~\bibnamefont {Kamp}}, \ and\ \bibinfo {author}
  {\bibfnamefont {S.}~\bibnamefont {H\"ofling}},\ }\href {\doibase
  10.1103/PhysRevB.92.205309} {\bibfield  {journal} {\bibinfo  {journal} {Phys.
  Rev. B}\ }\textbf {\bibinfo {volume} {92}},\ \bibinfo {pages} {205309}
  (\bibinfo {year} {2015}{\natexlab{b}})}\BibitemShut {NoStop}%
\bibitem [{\citenamefont {Konishi}\ \emph {et~al.}(2011)\citenamefont
  {Konishi}, \citenamefont {Nomura}, \citenamefont {Kumagai}, \citenamefont
  {Iwamoto}, \citenamefont {Arakawa},\ and\ \citenamefont
  {Kuwata-Gonokami}}]{Konishi2011}%
  \BibitemOpen
  \bibfield  {author} {\bibinfo {author} {\bibfnamefont {K.}~\bibnamefont
  {Konishi}}, \bibinfo {author} {\bibfnamefont {M.}~\bibnamefont {Nomura}},
  \bibinfo {author} {\bibfnamefont {N.}~\bibnamefont {Kumagai}}, \bibinfo
  {author} {\bibfnamefont {S.}~\bibnamefont {Iwamoto}}, \bibinfo {author}
  {\bibfnamefont {Y.}~\bibnamefont {Arakawa}}, \ and\ \bibinfo {author}
  {\bibfnamefont {M.}~\bibnamefont {Kuwata-Gonokami}},\ }\href {\doibase
  10.1103/PhysRevLett.106.057402} {\bibfield  {journal} {\bibinfo  {journal}
  {Phys. Rev. Lett.}\ }\textbf {\bibinfo {volume} {106}},\ \bibinfo {pages}
  {057402} (\bibinfo {year} {2011})}\BibitemShut {NoStop}%
\bibitem [{\citenamefont {Shitrit}\ \emph {et~al.}(2013)\citenamefont
  {Shitrit}, \citenamefont {Yulevich}, \citenamefont {Maguid}, \citenamefont
  {Ozeri}, \citenamefont {Veksler}, \citenamefont {Kleiner},\ and\
  \citenamefont {Hasman}}]{Shitrit2013}%
  \BibitemOpen
  \bibfield  {author} {\bibinfo {author} {\bibfnamefont {N.}~\bibnamefont
  {Shitrit}}, \bibinfo {author} {\bibfnamefont {I.}~\bibnamefont {Yulevich}},
  \bibinfo {author} {\bibfnamefont {E.}~\bibnamefont {Maguid}}, \bibinfo
  {author} {\bibfnamefont {D.}~\bibnamefont {Ozeri}}, \bibinfo {author}
  {\bibfnamefont {D.}~\bibnamefont {Veksler}}, \bibinfo {author} {\bibfnamefont
  {V.}~\bibnamefont {Kleiner}}, \ and\ \bibinfo {author} {\bibfnamefont
  {E.}~\bibnamefont {Hasman}},\ }\href {\doibase 10.1126/science.1234892}
  {\bibfield  {journal} {\bibinfo  {journal} {Science}\ }\textbf {\bibinfo
  {volume} {340}},\ \bibinfo {pages} {724} (\bibinfo {year} {2013})}\
  \BibitemShut
  {NoStop}%
  \bibitem [{\citenamefont {Rauter}\ \emph {et~al.}(2014)\citenamefont {Rauter},
  \citenamefont {Lin}, \citenamefont {Genevet}, \citenamefont {Khanna},
  \citenamefont {Lachab}, \citenamefont {Giles~Davies}, \citenamefont
  {Linfield},\ and\ \citenamefont {Capasso}}]{Rauter2014}%
  \BibitemOpen
  \bibfield  {author} {\bibinfo {author} {\bibfnamefont {P.}~\bibnamefont
  {Rauter}}, \bibinfo {author} {\bibfnamefont {J.}~\bibnamefont {Lin}},
  \bibinfo {author} {\bibfnamefont {P.}~\bibnamefont {Genevet}}, \bibinfo
  {author} {\bibfnamefont {S.~P.}\ \bibnamefont {Khanna}}, \bibinfo {author}
  {\bibfnamefont {M.}~\bibnamefont {Lachab}}, \bibinfo {author} {\bibfnamefont
  {A.}~\bibnamefont {Giles~Davies}}, \bibinfo {author} {\bibfnamefont {E.~H.}\
  \bibnamefont {Linfield}}, \ and\ \bibinfo {author} {\bibfnamefont
  {F.}~\bibnamefont {Capasso}},\ }\href {\doibase 10.1073/pnas.1421991112}
  {\bibfield  {journal} {\bibinfo  {journal} {Proceedings of the National
  Academy of Sciences}\ }\textbf {\bibinfo {volume} {111}},\ \bibinfo {pages}
  {E5623} (\bibinfo {year} {2014})}
   \BibitemShut {NoStop}%
\bibitem [{Note1()}]{Note1}%
  \BibitemOpen
  \bibinfo {note} {See a more detailed description of the cavity in the
  Supplementary material.}\BibitemShut {Stop}%
\bibitem [{\citenamefont {Konishi}\ \emph {et~al.}(2008)\citenamefont
  {Konishi}, \citenamefont {Bai}, \citenamefont {Meng}, \citenamefont
  {Karvinen}, \citenamefont {Turunen}, \citenamefont {Svirko},\ and\
  \citenamefont {Kuwata-Gonokami}}]{Konishi2008}%
  \BibitemOpen
  \bibfield  {author} {\bibinfo {author} {\bibfnamefont {K.}~\bibnamefont
  {Konishi}}, \bibinfo {author} {\bibfnamefont {B.}~\bibnamefont {Bai}},
  \bibinfo {author} {\bibfnamefont {X.}~\bibnamefont {Meng}}, \bibinfo {author}
  {\bibfnamefont {P.}~\bibnamefont {Karvinen}}, \bibinfo {author}
  {\bibfnamefont {J.}~\bibnamefont {Turunen}}, \bibinfo {author} {\bibfnamefont
  {Y.~P.}\ \bibnamefont {Svirko}}, \ and\ \bibinfo {author} {\bibfnamefont
  {M.}~\bibnamefont {Kuwata-Gonokami}},\ }\href {\doibase 10.1364/OE.16.007189}
  {\bibfield  {journal} {\bibinfo  {journal} {Opt. Express}\ }\textbf {\bibinfo
  {volume} {16}},\ \bibinfo {pages} {7189} (\bibinfo {year}
  {2008})}\BibitemShut {NoStop}%
\bibitem [{\citenamefont {Whittaker}\ and\ \citenamefont
  {Culshaw}(1999)}]{Whittaker1999}%
  \BibitemOpen
  \bibfield  {author} {\bibinfo {author} {\bibfnamefont {D.~M.}\ \bibnamefont
  {Whittaker}}\ and\ \bibinfo {author} {\bibfnamefont {I.~S.}\ \bibnamefont
  {Culshaw}},\ }\href@noop {} {\bibfield  {journal} {\bibinfo  {journal}
  {Phys.\ Rev.\ B}\ }\textbf {\bibinfo {volume} {60}},\ \bibinfo {pages} {2610}
  (\bibinfo {year} {1999})}\BibitemShut {NoStop}%
\end{thebibliography}
\end{document}